\documentclass[12pt]{iopart}
\usepackage{graphicx}
\begin{document}

\title{Lattice effects in the La$_{\rm 2-x}$Sr$_{\rm x}$CuO$_{\rm 4}$ compounds}
\author{E. Liarokapis$^{1,*}$, E. Siranidi$^{1}$, D. Lampakis$^{1}$, K. Conder$^{2}$, C. Panagopoulos$^{3}$}
\address{$^{1}$ Department of Physics, National Technical University, GR-15780 Athens, Greece}
\ead{$^{*}$eliaro@central.ntua.gr}
\address{$^{2}$ Laboratory for Solid State Physics, ETH Zurich, 8093 Zurich, Switzerland}
\address{$^{3}$ Cavendish Laboratory, University of Cambridge, J. J. Thomson Avenue,
Cambridge  CB3 0HE, United Kingdom and Department of Physics, University of Crete,
and Foundation for Research and Technology-Hellas, Heraklion, Crete, Greece}

\begin{abstract}

Systematic Raman studies on several cuprates (YBa$_{\rm
2}$Cu$_{\rm 3}$O$_{\rm x}$, YBa$_{\rm 2}$Cu$_{\rm 4}$O$_{\rm 8}$
or Bi$_{\rm 2}$Sr$_{\rm 2}$CaCu$_{\rm 2}$O$_{\rm 8}$) have shown
that at optimal doping the compounds are at the edge of lattice
instability; once this level is exceeded, by means of doping or
applying external hydrostatic pressure, the changes in the
transition temperature are accompanied by spectral modifications.
There are strong indications that the reduction in T$_{\rm c}$ is
correlated with a separation into nanoscale phases, which involve
the oxygen atoms of the CuO$_{\rm 2}$ planes. In this work,
modifications with doping in the Raman spectra of the La$_{\rm
2-x}$Sr$_{\rm x}$CuO$_{\rm 4}$ compound are presented, which show
that spin or charge ordering is coupled with lattice distortions
in the whole doping region.

\end{abstract}

\pacs{PACS numbers:  }

\noindent{\it Keywords\/}: Cuprates; Raman scattering; Phase
separation

\submitto{\JPCM} \maketitle

\section{Introduction}

More than two decades after the discovery of high temperature
superconductors (HTSCs) \cite{Bednorz} the role of the lattice in
the pairing mechanism of those compounds is still unclear, with
many conflicting experiments. Raman spectroscopy is a powerful
technique that can provide direct evidence about small lattice
distortions. Although it probes only the q$\approx$0 phonons in
the Brillouin zone, due to its extreme sensitivity, it can detect
a weak interaction of the lattice with the carriers or even the
spin ordering. In several cuprates (YBa$_{\rm 2}$Cu$_{\rm
3}$O$_{\rm x}$, YBa$_{\rm 2}$Cu$_{\rm 4}$O$_{\rm 8}$, or Bi$_{\rm
2}$Sr$_{\rm 2}$CaCu$_{\rm 2}$O$_{\rm 8}$) it was found that at
optimal doping the compounds are at the edge of lattice
instability, which once exceeded by doping or external hydrostatic
pressure the T$_{\rm c}$ reduction is accompanied by spectral
modifications in the phonons that involve vibrations of the plane
oxygen (O$_{\rm pl}$) atoms \cite{Kaldis,Lampakis}. This mode
softens suddenly above optimal doping or it appears as a double
peak under hydrostatic pressures \cite{Kaldis,Lampakis}. By
slightly tilting the polarization axis, it has been shown that the
relative intensity of the two peaks varies, clearly indicating
that the new peak corresponds to a tilting of the octahedra
\cite{Lampakis}. This behavior seems to be quite general in all
cuprates studied up to now and it was attributed to a (nano)scale
phase separation of a compound that seems to be close to an
instability.

Another typical example of the cuprate superconducting family that
shows a variety of phases with doping and temperature
\cite{Keimer,Johnston} is La$_{\rm 2-x}$Sr$_{\rm x}$CuO$_{\rm 4}$.
This compound displays a characteristic soft mode at $\approx$120
cm$^{\rm -1}$ (for x=0), that has the A$_{\rm g}$ symmetry and it
is attributed to the tilting vibrations of the CuO$_{\rm 6}$
octahedra about the diagonal (110) axis \cite{Sugai,Weber}. The
mode is activated in the Low Temperature Orthorhombic (LTO) phase
disappearing as we approach the High Temperature Tetragonal (HTT)
more symmetric phase \cite{Sugai,Weber,Burns}. From the rest of
the Raman active phonons, the two strong modes are of A$_{\rm g}$
symmetry and are attributed to the vibrations of the La/Sr and the
apex oxygen atoms along the c axis \cite{Sugai,Weber}. Other
weaker modes appear by Sr doping mostly at low temperatures and
for doping levels for which the compound is superconducting. The
peculiar characteristics of the soft mode and the origin of the
weaker ones will be examined below.

\section{Experimental}

Individual microcrystallites from a selected series of high
quality La$_{\rm 2-x}$Sr$_{\rm x}$Cu$^{\rm 16,18}$O$_{\rm 4}$
(85\% substitution $^{\rm 16}$O by $^{\rm 18}$O) polycrystalline
compounds, with Sr doping in the range 0.00$\leq$x$\leq$0.3, have
been studied in the 7.5-350K temperature region using Raman
spectroscopy. The Raman spectra were obtained in the approximate
y(zz)y and y(xx)y (or x(yy)x) scattering configurations or a
mixture of them as the x and y axes could not be discriminated in
the twinned samples. The triple spectrometer Jobin-Yvon T64000 was
used, which is equipped with a liquid-nitrogen-cooled
charge-coupled-device (CCD) and a microscope ($\times$100
magnification). Low temperatures were achieved by using an
open-cycle Oxford cryostat operating either with liquid nitrogen
or liquid helium. The 514.5nm wavelength of an Ar$^{\rm +}$ laser
was used for excitation at a power level of 0.05-0.1
mW/$\mu$m$^{\rm 2}$. It appeared that the polycrystalline samples
under vacuum were very sensitive to the local heating and special
care was necessary to secure that the laser power will be low
enough to avoid any local heating during the many hours
measurements. Based on the soft mode energy and width we have
estimated that the local heating of the sample due to the laser
beam was less than 10K. As a result, accumulation times for each
measurement were of the order of 4-5 hours.

\section{Results and discussion}

Typical Raman spectra of the LSCO compound for selected doping and
temperatures for the parallel polarizations of the incoming and
scattering light along the c-axis (zz spectra) and the ab planes
(xx spectra) are presented in Figs.1a \& b respectively. The data
from two oxygen isotopes ($^{\rm 16}$O and $^{\rm 18}$O) are
presented in order to identify the modes related with vibrations
of the oxygen atoms.

In Fig.1a, all five A$_{\rm g}$ phonons can be seen. Three of
these, the La/Sr, the apex oxygen, and the weak mode at
$\approx$273 cm$^{\rm -1}$ involve ion vibrations along the c-axis
and the rest of the Raman active modes (the soft mode and the
other very weak mode at $\approx$156 cm$^{\rm -1}$) are attributed
to the tilting vibrations of the octahedra \cite{Weber}. The
energy of the soft mode decreases and its width increases rapidly
with increasing temperature as expected \cite{Sugai}. In the range
T=285-305 K for x=0.0 and 77-100K for x=0.015 both energy and
width show an abnormal behavior (Figs.2a \& b respectively). In
this region, the energy of the soft mode remains constant and
independent of temperature, while the width increases
substantially. With the further increase in temperature the energy
of the soft mode decreases again. The effect in the La$_{\rm
2}$CuO$_{\rm 4}$ compound (Fig.2a) is much more pronounced and it
occurs at higher temperatures than for the other Sr concentration
(Fig.2b). The results from the oxygen isotopic substituted
compounds show that there is no isotope effect on the temperature
ranges where this abnormal behavior is observed.

According to the phase diagram of LSCO (Fig.3), the structural
LTO$\rightarrow$HTT transition temperature, for the insulating
La$_{\rm 2}$CuO$_{\rm 4}$ sample is expected at $\approx$560K and
cannot be related with the abnormal behavior observed in the range
T=285-305K. As Fig.3 schematically shows, close to room
temperature there is a N\'{e}el transition to the
antiferromagnetic phase \cite{Keimer,Sugai2} for x=0.0, while for
x=0.015 the N\'{e}el temperature is less than 100K falling to zero
for x$\approx$0.02 \cite{Keimer}. Both temperatures agree with the
ranges where the abnormal behavior of the soft mode was observed
in the two concentrations. Besides, the N\'{e}el temperature is
not affected by the oxygen isotopic substitution \cite{Zhao} and
therefore any effect related with the spin ordering should be
independent of the isotopic substitution. The data in Fig.2 seem
to support the idea that the effects observed at T=285-305K for
x=0 and 77-100K for x=0.015 are mostly related with the
antiferromagnetic (AF) ordering. The present data provide strong
evidence for the coupling of the antiferromagnetic ordering with
the lattice at least for the low doping levels. Such
magnetoelastic coupling between spin ordering and the lattice
(octahedra of the oxygen atoms) has been proposed in
Ref.\cite{Neto}. Here there is a clear proof that the tilting
(soft) mode of the octahedra is affected by the spin ordering.
With increasing doping the LTO$\rightarrow$HTT structural phase
transition temperature decreases to disappear for x$\simeq$0.22.
The soft mode will also disappear at this doping level. On the
other hand, the AF ordering exists only for low doping levels. It
is therefore interesting to investigate the coupling of the soft
mode in the whole doping region. Our data indicate that there is
no such anomaly for higher Sr concentrations, supporting our
assumption that the effect observed (Fig.2) is due to the coupling
of the lattice (tilting of the octahedra) with the spin ordering.

In the y(xx)y (or x(yy)x) polarization Raman spectra (Fig.1b)
additionally to the A$_{\rm g}$-symmetry phonons at $\approx$229
cm$^{\rm -1}$ of the La/Sr ions and at $\approx$429 cm$^{\rm -1}$
of the apical oxygen, certain new bands at $\approx$150 cm$^{\rm
-1}$, $\approx$280 cm$^{\rm -1}$, and $\approx$370 cm$^{\rm -1}$
appear mainly at low temperatures and in the superconducting
doping region 0.03$<$x$<$0.27. All bands, including the new ones,
disappear in the crossed polarization and therefore they
approximately have the A$_{\rm 1}$-symmetry. This apparent
symmetry excludes the possibility for the three new bands to be
associated with the B$_{\rm 1g}$, B$_{\rm 2g}$, and B$_{\rm 3g}$
Raman active phonons. The study of the effect of the oxygen
isotopic substitution on the three new bands proves that the band
at $\approx$150 cm$^{\rm -1}$ has a $\approx$(1.5$\pm$0.5)\%
energy shift, the band at $\approx$280 cm$^{\rm -1}$ a
$\approx$(3.5$\pm$0.5)\% energy shift, while the band at
$\approx$370 cm$^{\rm -1}$ follows very well the mass harmonic law
of a purely oxygen mode ($\approx$5.0\%) (Fig.4).

In previous works \cite{Sugai,Sugai2,Sugai3} the bands at
$\approx$150 cm$^{\rm -1}$ and $\approx$370 cm$^{\rm -1}$ have
been attributed to the TO IR modes with eigenvectors along the
CuO$_{\rm 2}$ planes, while there was no proposed association for
the $\approx$280 cm$^{\rm -1}$ band. Even though we agree that the
new bands which appear in the xx polarization Raman spectra are IR
modes, based on our experimental data from the oxygen isotopic
substitution, we believe that the assignment given in the work of
Refs.\cite{Sugai,Sugai2,Sugai3} is not correct. From the
association of the D$_{\rm 2h}$ point group (for the orthorhombic
phase) with its subgroups \cite{Rousseau} one can find that with
the loss of inversion symmetry (which is necessary for IR modes to
become Raman active), the B$_{\rm 1u}$, B$_{\rm 2u}$, B$_{\rm
3u}$, and A$_{\rm u}$ phonons of D$_{\rm 2h}$ are associated with
the A$_{\rm 1}$-symmetry modes of the $\rm C^{2v}_z$, $\rm
C^{2v}_y$, $\rm C^{2v}_x$, and D$_{\rm 2}$ subgroups respectively.
Comparing the observed peaks with IR-active phonons of similar
energy \cite{Bazhenov} and concerning the results from the oxygen
isotopic substitution certain possibilities are revealed. The band
at $\approx$150 cm$^{\rm -1}$ can be related with the following IR
modes; the B$_{\rm 3u}$ of La/Sr (following the Abma notation of
the orthorhombic phase), the B$_{\rm 1u}$ of Cu$_{\rm pl}$, the
B$_{\rm 1u}$ or the silent A$_{\rm u}$ of the La/Sr atoms and
finally the B$_{\rm 2u}$ of the O$_{\rm ab}$ with a mixture of Cu
and La/Sr atoms. The wide band at $\approx$280 cm$^{\rm -1}$ with
one of the four phonons; the B$_{\rm 1u}$-symmetry of O$_{\rm
apex}$ and O$_{\rm ab}$, the silent A$_{\rm u}$ of the O$_{\rm
ab}$ atoms and the B$_{\rm 3u}$ phonon of the Cu$_{\rm pl}$ or of
the Cu with a mixture of La/Sr and O$_{\rm ab}$ atoms. Finally,
the $\approx$370 cm$^{\rm -1}$ with the B$_{\rm 3u}$ phonon of the
O$_{\rm apex}$.

It is quite possible that the $\approx$150 cm$^{\rm -1}$ and
$\approx$280 cm$^{\rm -1}$ peaks are composite bands consisting of
several modes, as our assignment has shown. In fact, the band at
$\approx$150 cm$^{\rm -1}$ appears in certain concentrations and
temperatures as multiple (two or three) peaks. Similarly, the
$\approx$280 cm$^{\rm -1}$ band is very wide (Fig.1b) to be
considered as a single mode. Besides, in the compounds with oxygen
isotopic substitution, the $\approx$280 cm$^{\rm -1}$ band appears
with a modified width, which could not happen unless it was
consisting of multiple modes, some of them related (and shifting)
with vibrations of the oxygen atoms and others involving motion of
the Cu$_{\rm pl}$ or the La/Sr atoms.

Since the IR modes appear in the Raman spectra for those doping
levels where the compound becomes superconducting and mainly at
low temperatures, it is important to investigate possible reasons
for the inversion symmetry breaking. Any connection of the effect
with the orthorhombic to tetragonal phase transition can be
excluded, because there are no IR modes at ambient temperature in
the pure La$_{\rm 2}$CuO$_{\rm 4}$ compound, which is
orthorhombic. The modes appear with Sr doping and increase in
intensity showing a maximum around optimal doping and disappear in
the overdoped region. One could attribute them to the disorder
introduced by the Sr substitution. Such effect should increase
gradually with the amount of Sr and not disappear with
x$\approx$0.25 as the IR modes. Other q$\cong$0 Raman active modes
can be also excluded, since the new modes persist even in the
tetragonal phase, where very few phonons can be observed. One
cannot exclude phonons from the edges of the Brillouin zone. Such
mode will indicate the formation of superstructures along some
direction, which could be the case. The other possibility is just
the breaking of the local inversion symmetry. This can happen from
the formation of stripes or polarons. Stripes will probably induce
a superstructure, so it could also activate phonons from other
high symmetry points of the Brillouin zone. The formation of
polarons or stripes could justify the connection of the new modes
and the inversion symmetry breaking with the appearance of
superconductivity.

\section{Conclusions}

For several cuprates, modifications in the Raman spectra with
chemical doping (or pressure) have been observed, with new modes
appearing or others split. These changes are correlated with
modifications in T$_{\rm c}$ and point to a separation and
coexistence of phases. For La$_{\rm 2-x}$Sr$_{\rm x}$CuO$_{\rm 4}$
modifications have been observed in the soft mode energy for very
low doping levels for temperatures that coincide with the
antiferromagnetic ordering. In this way, the soft mode seems to be
coupled with spin ordering. Besides, new modes appear at doping
levels where the compound is superconducting at temperatures above
T$_{\rm c}$ apparently from a local symmetry breaking of the
lattice. These changes can be attributed to the formation of
stripes or polarons, that reduce the local symmetry and are
related with the superconductivity.

\ack Work supported through the project E.C. STREP No 517039
project "COMEPHS". The authors want to thank P.Auban-Senzier and
C. Pasquier for the resistivity experiments.

\bigskip

\section*{FIGURE CAPTIONS}

Fig. 1. (a) Typical spectra for selected temperatures and doping
levels of the La$_{\rm 2-x}$Sr$_{\rm x}$Cu$^{\rm 16,18}$O$_{\rm
4}$ compounds in the zz polarization. (b) Typical spectra for
selected temperatures of the La$_{\rm 2-x}$Sr$_{\rm x}$Cu$^{\rm
16}$O$_{\rm 4}$ compound in the xx polarization.

\noindent Fig. 2. (a) The temperature dependence of the energy and
width of the soft mode for the La$_{\rm 2}$Cu$^{\rm 18}$O$_{\rm
4}$ compound. (b) The temperature dependence of the energy and
width of the soft mode for the La$_{\rm 1.985}$Sr$_{\rm
0.015}$Cu$^{\rm 16}$O$_{\rm 4}$ compound.

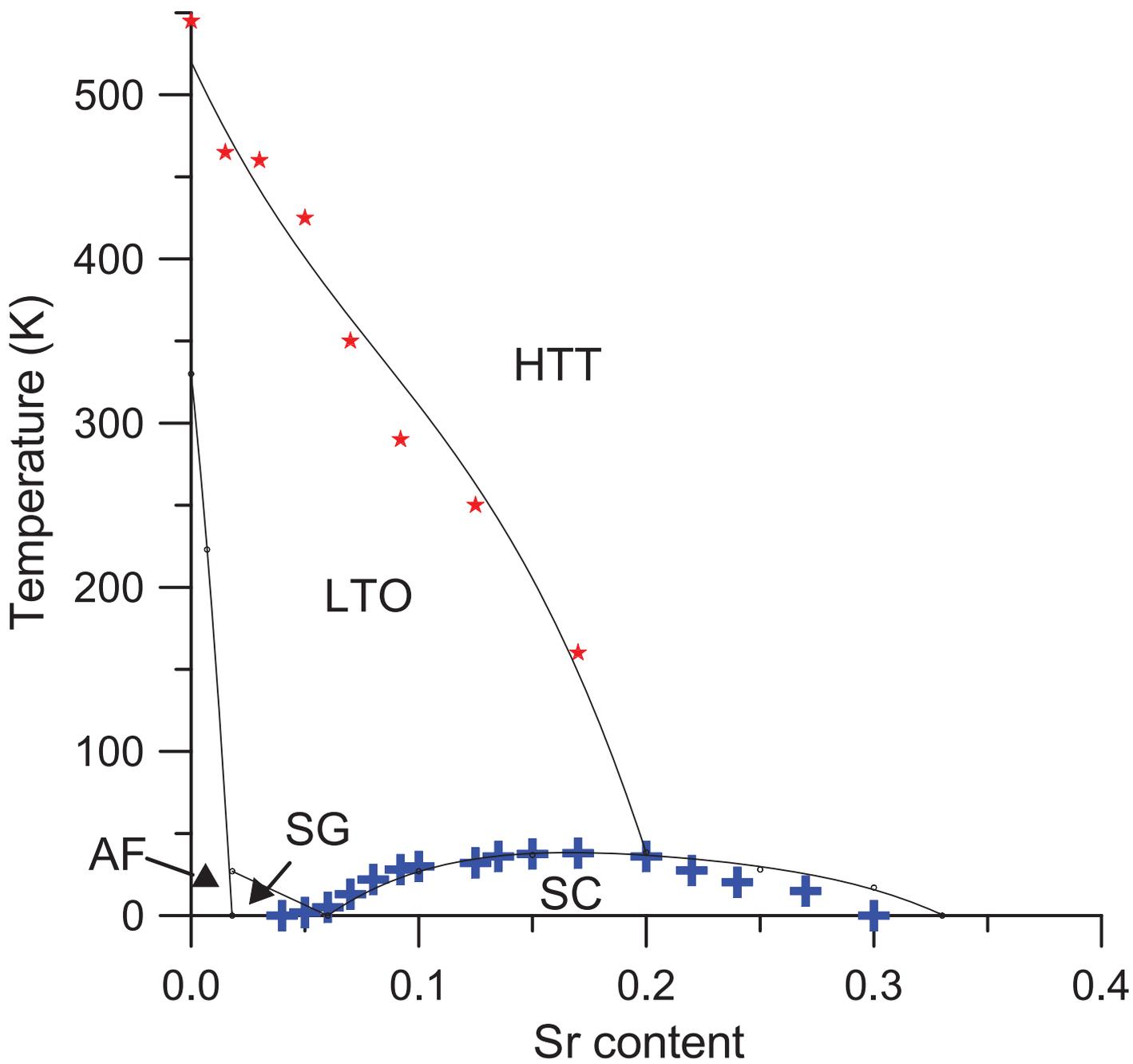
\noindent Fig. 3. Phase diagram of LSCO. Daggers and stars
indicate experimental data points.

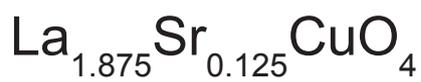
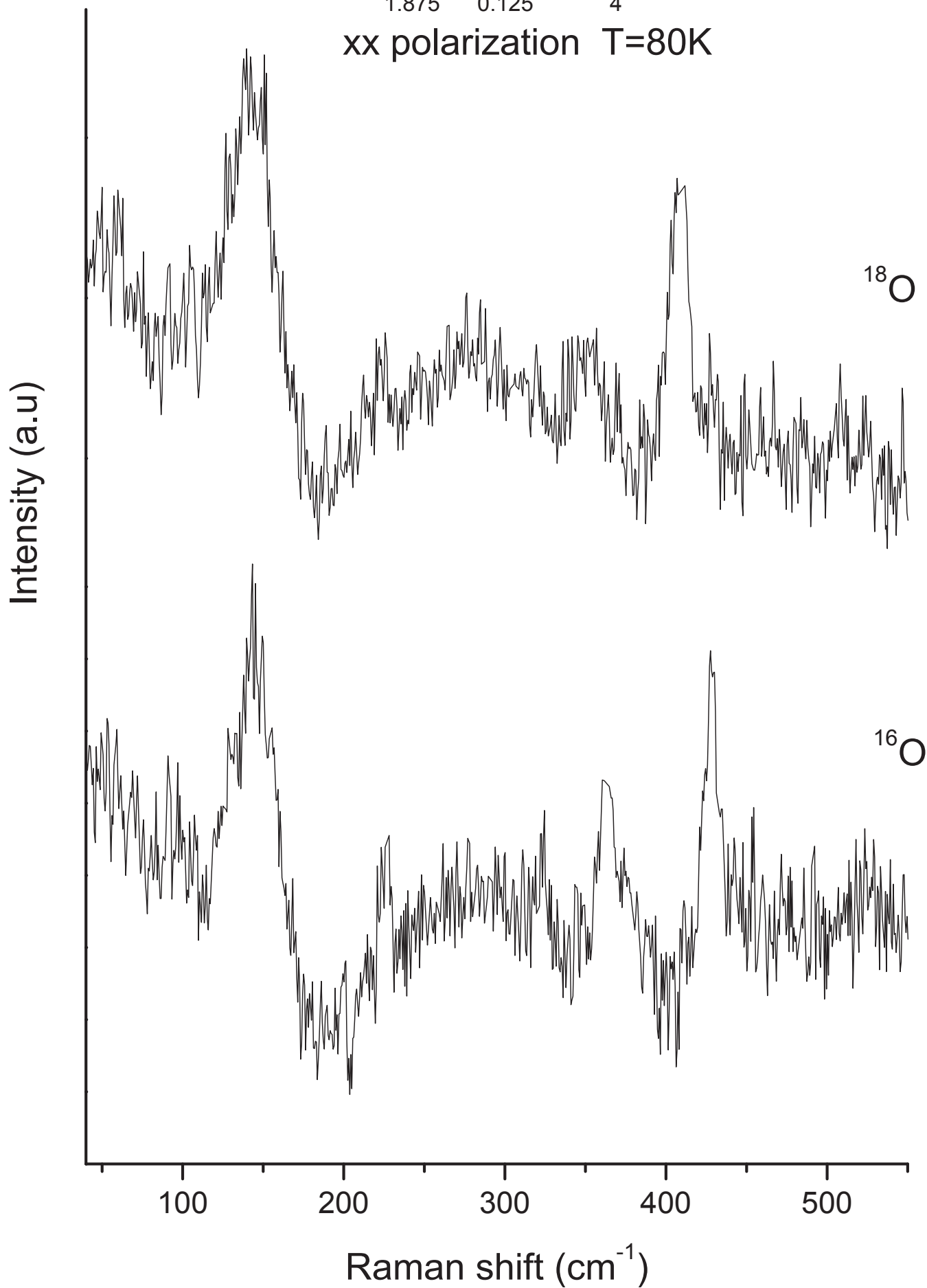
\noindent Fig. 4. Raman spectra for the parallel polarizations of
the two oxygen isotopes of the La$_{\rm 1.875}$Sr$_{\rm
0.125}$CuO$_{\rm 4}$ compound.

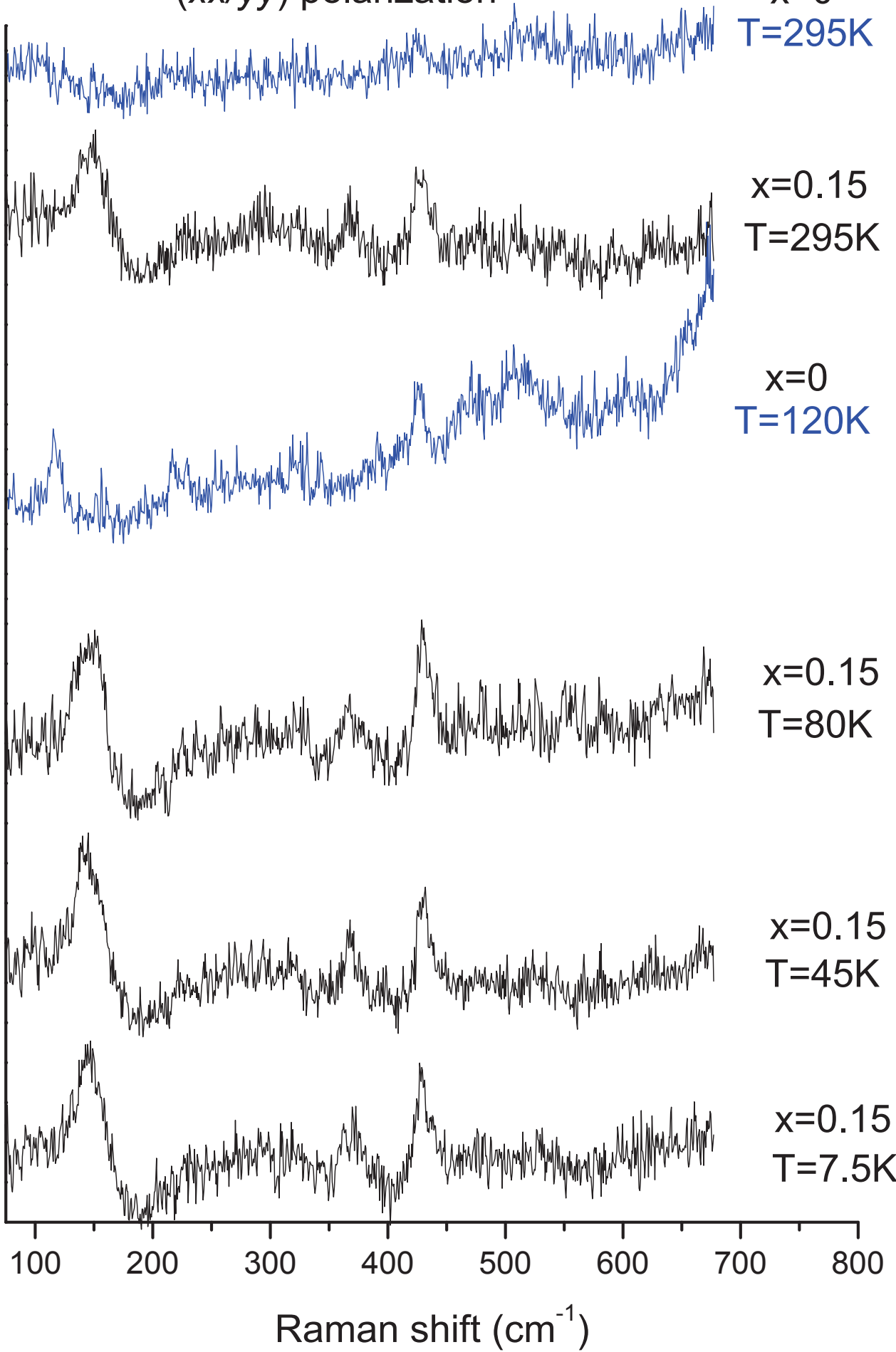
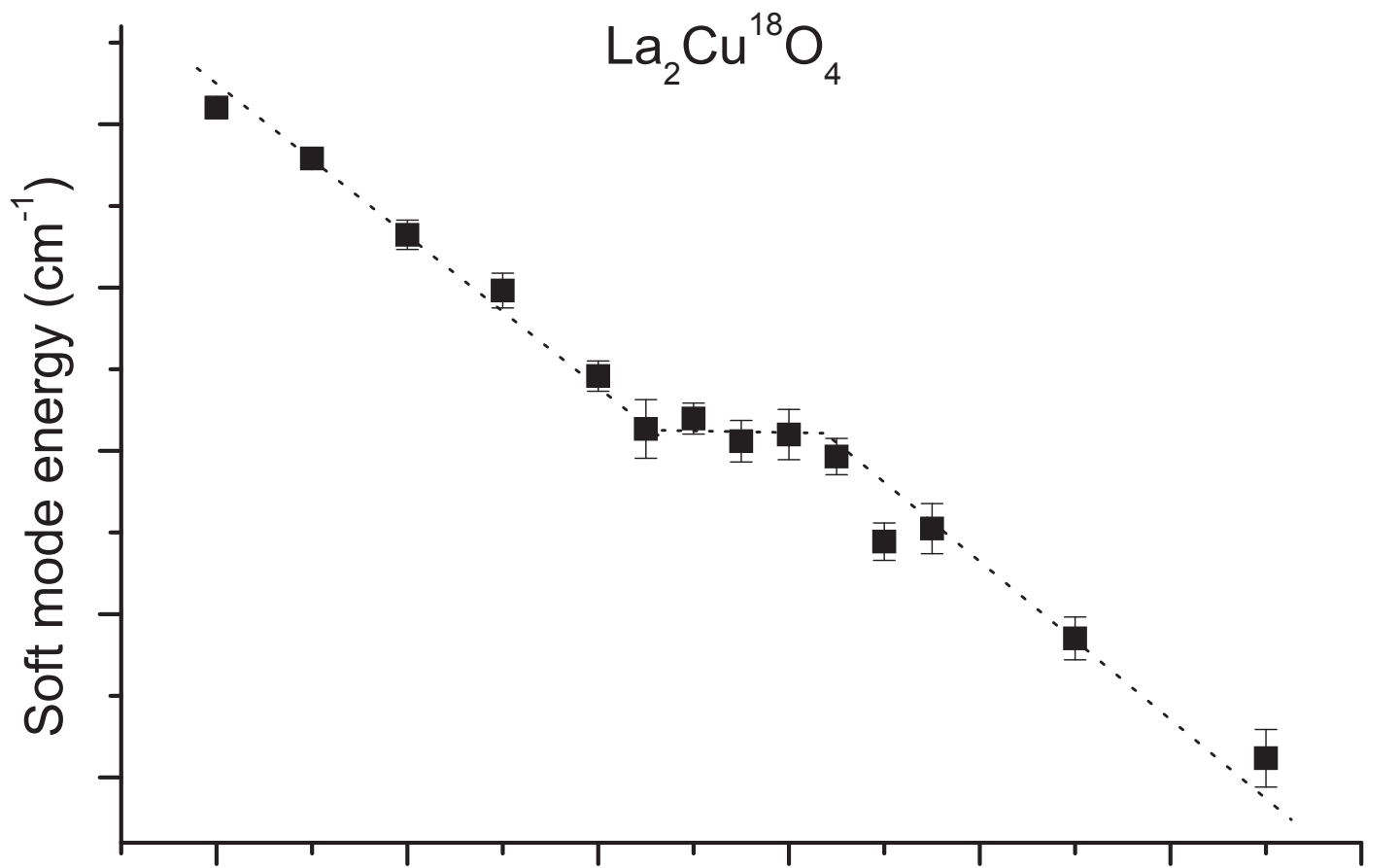
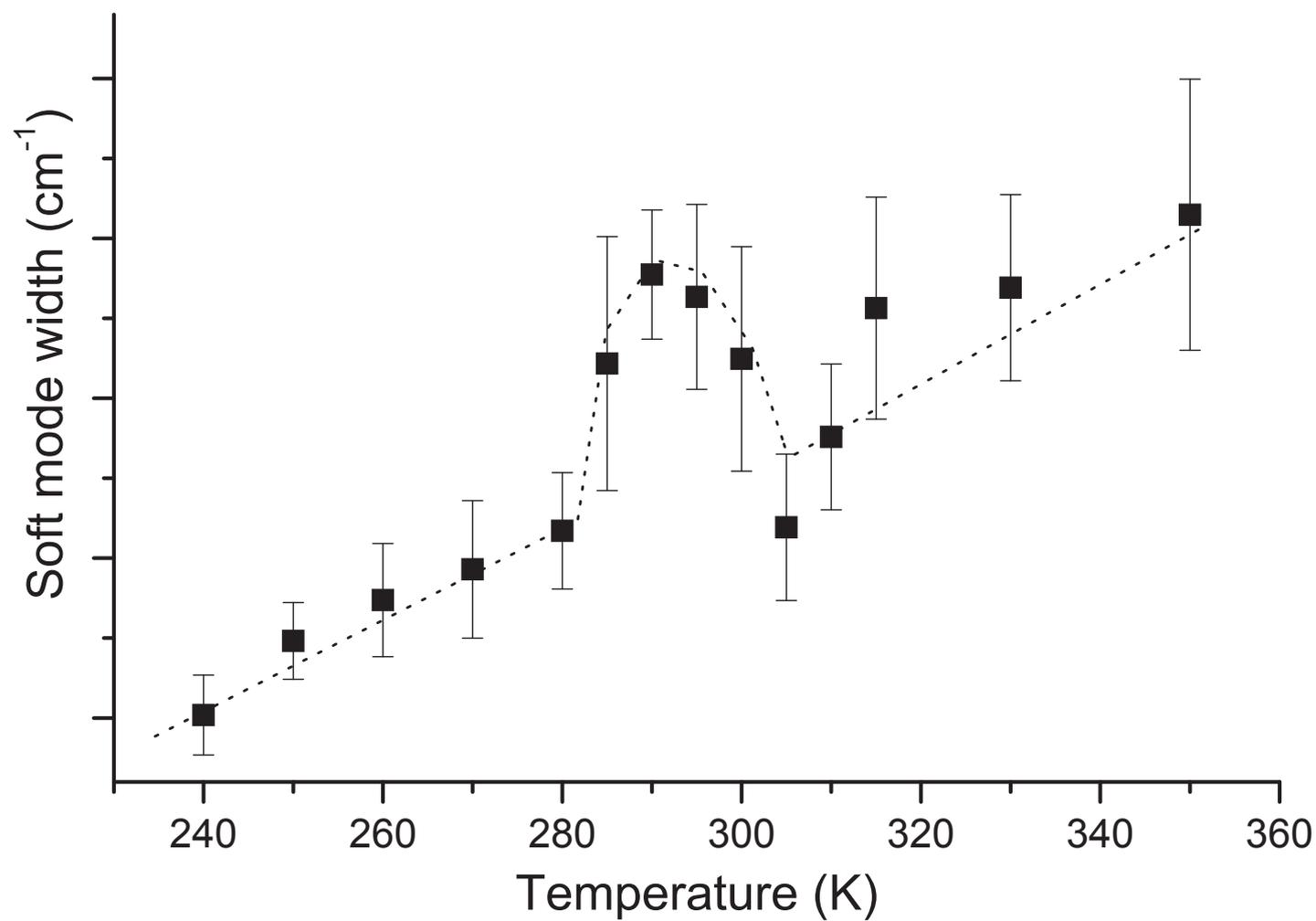
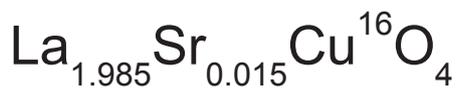
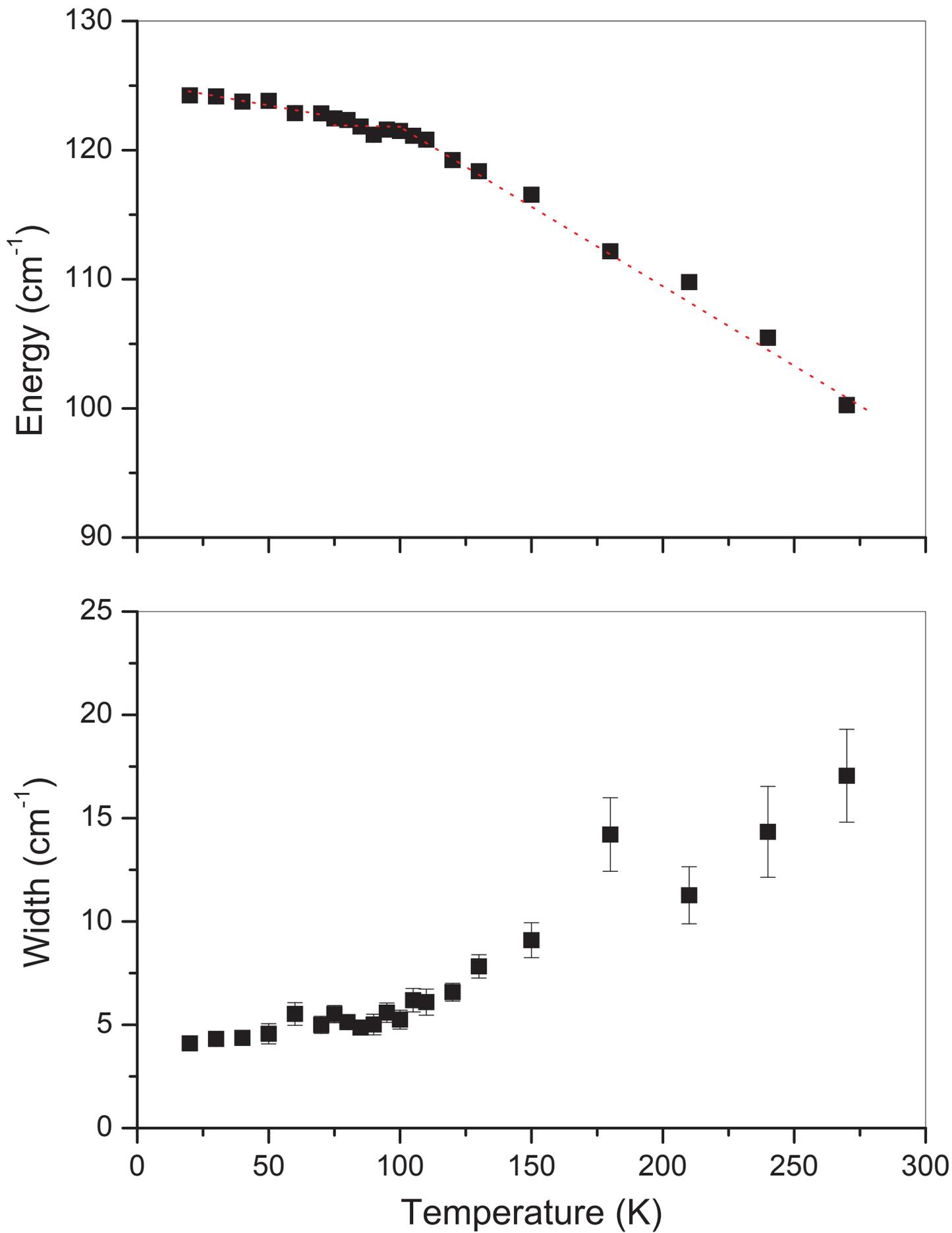
\end{document}